\documentclass[preprint,aps,prb,onecolumn,groupedaddress,superscriptaddress,nobibnotes]{revtex4-2}

\usepackage{amsmath}
\usepackage{amssymb}
\usepackage{graphicx}
\usepackage{dcolumn}
\usepackage{bm}
\usepackage{color}
\usepackage{subfig}
\usepackage[normalem]{ulem}
\usepackage[none]{hyphenat}
\usepackage{soul}
\usepackage{hyperref}
\usepackage{lineno}
\usepackage{titlesec}
\titleformat{\section}{\normalfont\large\bfseries}{\thesection}{1em}{}

\newcommand{\BT}{Bi$_2$Te$_3$}
\newcommand*{\BS}{Bi$_2$Se$_3$}

\newcommand*{\MBS}{MnBi$_2$Se$_4$}
\newcommand*{\MBT}{MnBi$_2$Te$_4$}
\newcommand*{\MBTBT}{MnBi$_4$Te$_7$}
\newcommand*{\MBTBTBT}{MnBi$_6$Te$_{10}$}
\newcommand*{\MST}{MnSb$_2$Te$_4$}

\newcommand*{\MBST}{Mn(Bi$_{1-x}$Sb$_x$)$_2$Te$_4$}

\newcommand*{\BST}{(Bi$_x$Sb$_{1-x}$)$_{2}$Te$_3$}

\newcommand{\PreserveBackslash}[1]{\let\temp=\\#1\let\\=\temp}

\begin{document}

\title{Intrinsic magnetic topological insulators of the \MBT\, family}

\author{Alexandra Yu.~Vyazovskaya}
\affiliation{Tomsk State University, 634050 Tomsk, Russia}
\affiliation{Saint Petersburg State University, 199034 Saint Petersburg, Russia}

\author{Mihovil Bosnar}
\affiliation{Donostia International Physics Center, 20018 Donostia-San Sebasti\'an, Spain}
\affiliation{Departamento de Pol\'imeros y Materiales Avanzados: F\'isica, Qu\'imica y Tecnolog\'ia, Facultad de Ciencias Qu\'imicas, Universidad del Pa\'is Vasco UPV/EHU, 20018 Donostia-San Sebasti\'an, Spain}
\affiliation{{Present address: Department of Physics, University of Zagreb, Bijenička cesta 32, 10000 Zagreb}}

\author{Evgueni V.~Chulkov}
\email{evguenivladimirovich.tchoulkov@ehu.eus}
\affiliation{Donostia International Physics Center, 20018 Donostia-San Sebasti\'an, Spain}
\affiliation{Centro de Física de Materiales (CFM-MPC), Centro Mixto (CSIC-UPV/EHU), 20018 Donostia-San Sebasti\'an, Spain}
\affiliation{Saint Petersburg State University, 199034 Saint Petersburg, Russia}

\author{Mikhail M.~Otrokov}
\email{mikhail.otrokov@csic.es}
\affiliation{Instituto de Nanociencia y Materiales de Aragón (INMA), CSIC-Universidad de Zaragoza, 50009 Zaragoza, Spain}

\begin{abstract}

This short review appears on the occasion of the fifth anniversary of discovery of intrinsic magnetic topological insulators (MTIs) of the \MBT\, family, which have attracted a great deal of attention recently. This family of materials has been discovered in attempts to increase the observation temperature of the quantum anomalous Hall effect as well as to facilitate the eventual realization of the topological magnetoelectric effect. Therefore, we first briefly introduce these effects, then describe the experimental state-of-the-art in the MTIs field just prior to \MBT\, appearance, after which we discuss the basic properties of this material and its family. Finally, we overview the exciting progress made during five years of intense research in this field.
\end{abstract}

\date{\today}

\maketitle

{{\section*{Introduction}}

Introducing magnetism in a nonmagnetic topological insulator (TI) is known to give rise to fascinating and technologically promising phenomena \cite{Hasan2010, Qi.rmp2011}. One of them is the quantum anomalous Hall effect (QAHE) \cite{Haldane.prl1988, Liu.prl2008, Yu.sci2010, Chang.rmp2023}, which can be observed in the magnetic thin films featuring an inverted band gap due to spin-orbit coupling, i.e. two-dimensional magnetic TIs (MTIs). The bulk-edge correspondence\cite{Shapiro.review} dictates that at the physical border of such materials there must be a gapless topological edge state. Remarkably, an electron moving in such a state at any given edge of the sample can propagate only in one direction and cannot backscatter. Such a chirality of the edge state makes the QAHE attractive for the dissipationless transport applications. For example, it can be used to create interconnect devices that electrically connect the components of an integrated circuit \cite{Zhang.patent2015}.
The experimental hallmark of the QAHE is a vanishing longitudinal conductivity $\sigma_{xx}$ along with a transversal conductivity $\sigma_{xy}$ quantized to integer multiples of the conductance quantum, $C$e$^2/h$ (for the resistivities, $\rho_{xx}=0$ and $\rho_{yx}=h/(C$e$^2)$). Here, e is the electron charge, $h$ is the Planck’s constant, and $C$ is a dimensionless integer called the first Chern number, which is a topological invariant for these kind of systems \cite{Hasan2010, Qi.rmp2011, Chang.rmp2023}.

Another fundamental phenomenon arising in MTIs is the topological magnetoelectric effect (TME) \cite{Qi.prb2008, Qi.rmp2011,Mahon.prr2024}. When the TI surfaces are gapped due to magnetism, it should respond to the application of an external electric (magnetic) field by generation of a (an) magnetic (electric) polarization that appears to be quantized. An intriguing TME-related phenomenon is the solid-state embodiment of the axion electrodynamics \cite{Qi.rmp2011, Essin.prl2009, Sekine.jap2021}, arising from a peculiar analogy with a field theory, where the axion field is said to generate electrodynamics with exactly the same Lagrangian that describes the TME in TIs. A material possessing these exotic properties is called axion insulator. Other interesting implications of the unique magnetolectric properties of MTIs are the quantized Kerr and Faraday effects \cite{Qi.rmp2011, Tse.prl2010}.

The observation of the QAHE in the Cr-doped TI \BST\, \cite{Chang.sci2013, Checkelsky.natp2014} gave a strong impetus to a worldwide study of these systems. Nowadays, magnetically-doped TIs reproducibly show very robust high-precision QAHE at mK temperatures and they are currently being considered for metrological applications \cite{Okazaki.nphys2022}. Besides, the QAHE is achieved not only in the MTIs, but also in the twisted bilayer graphene \cite{Serlin.sci2020} and transition metal dichalcogenides heterostructures \cite{Li.nat2021}. Measuring TME in a magnetic TI proved more challenging, but it was eventually achieved, too \cite{Mogi.nphys2022}. However, a random distribution of the magnetic atoms in the magnetically-doped TIs leads to strongly inhomogeneous magnetic and electronic properties of these materials \cite{Lee.pnas2015, Lachman.sciadv2015,  Krieger.prb2017}, restricting the observation of these effects to very low temperatures \cite{Mogi.apl2015}. More specifically, the atomic disorder leads to the fluctuation of both the size and energy position of the Dirac point gap   across the surface of these materials, as imaged with scanning tunneling spectroscopy in \cite{Lee.pnas2015}. As a consequence, this gap has not been observed in angle-resolved photoemission spectroscopy (ARPES) for Cr- or V-doped \BST\, with stoichiometries close
to those showing QAHE \cite{Li.srep2016, Kim.jap2021}.

\begin{figure}
	\includegraphics[width=0.4\textwidth]{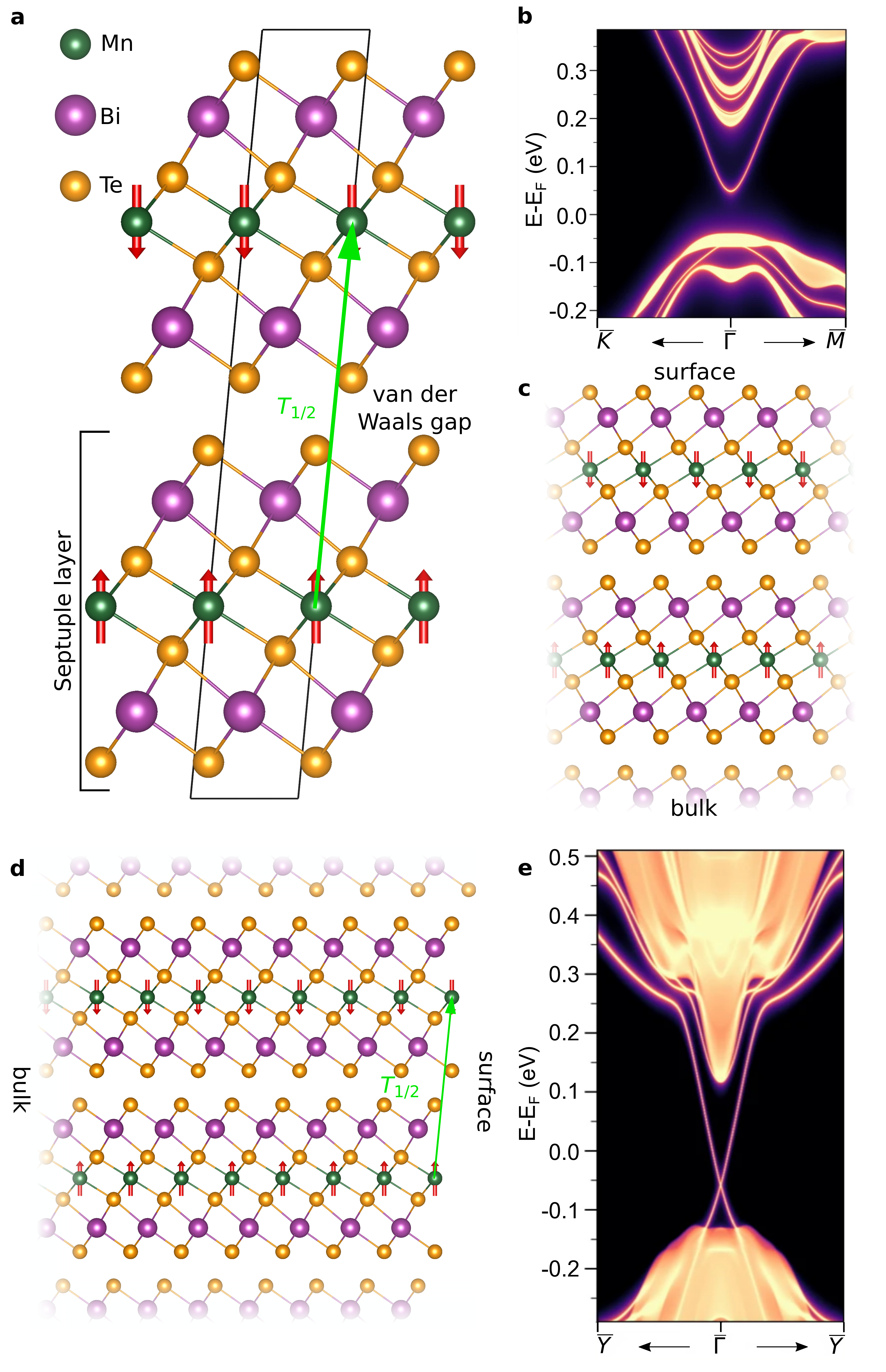}
	\caption{{\bf{Crystal and magnetic structures as well as the calculated low-energy surface spectra of \MBT.}} {\bf a} Crystal and magnetic structures of bulk \MBT. Red arrows denote Mn local moments, while the $T_{1/2}$ translation is shown in green. {\bf b, e} Surface electronic band structures of \MBT\, calculated for the (0001) ($S$-breaking; {\bf c}) and for the $(10\overline{1}1)$ ($S$-preserving; {\bf d}) surface terminations, respectively, using the \emph{ab-initio}-based tight-binding approach. The regions with a continuous spectrum correspond to the three-dimensional bulk states projected onto a two-dimensional Brillouin zone. Observe the gapped (gapless) character of the $S$-breaking ($S$-preserving) surface. {{\bf b}, {\bf e} are reproduced with permission from
    Ref. \cite{Otrokov.nat2019}, Copyright Springer-Nature, 2019.}}
	\label{fig1}
\end{figure}
As an alternative to the magnetically doped TIs, a new class of magnetic topological matter emerged: \emph{intrinsic} MTI compounds, the first representative of this class being antiferromagnetic (AFM) van der Waals material \MBT\, \cite{Otrokov.nat2019, Zhang.prl2019, Li.sciadv2019, Gong.cpl2019, Otrokov.prl2019, Rienks.nat2019}. This discovery was preceded by both theoretical and experimental studies of the magnetic topological heterostructures such as \MBT/\BT(0001), \MBS/\BS(0001) and others that were based on the thin films of Mn(Bi/Sb)$_2$(Te/Se)$_4$ \cite{Otrokov.2dmat2017, Otrokov.jetpl2017, Hirahara.nl2017, Eremeev.jac2017, Hagmann.njp2017, Eremeev.nl2018}.
It is necessary to say that the AFM TI state of matter was for the first time theoretically proposed in 2010 by Mong, Essin, and Moore \cite{Mong.prb2010}. They introduced a ${\mathbb{Z}}_{2}$ topological classification of AFM insulators, provided by the $S=\Theta T_{1/2}$-symmetry, where $\Theta$ is time-reversal and $T_{1/2}$ is the primitive lattice translation. However, during almost a decade after the prediction no material satisfying these conditions was observed experimentally.

{In this short review, we provide the current state-of-the-art in the research field that has emerged following the discovery of \MBT. Our aim is to offer a concise yet comprehensive overview that serves both as a quick update for researchers in the field and an accessible entry point for newcomers. We first introduce the basic properties of \MBT, followed by a discussion of the key topics, including the sample dependence of the gap in its topological surface state, native defects and their possible impact on the gap, and the surface electronic structure of \MBT\ above the Néel temperature. This is followed by an exploration of the intrinsic MTIs of the \MBT\ family and, eventually, the unique properties of \MBT\ in the two-dimensional limit. Throughout the review, we examine the most significant and recent advancements in the field, while analyzing and highlighting its key challenges. Finally, in the outlook, we briefly discuss the exciting opportunities that the \MBT\ family of compounds presents for future research and applications.}

{{\section*{Basic properties of {\MBT}}}

The first reference to \MBT\, dates back to 2013 when it was synthesized in the powder form and its structure as well as thermoelectric properties were studied \cite{Lee.cec2013}.
It crystallizes in the trigonal $R\bar 3m$-group structure~\cite{Lee.cec2013, Aliev.jac2019, Zeugner.cm2019} comprising septuple layer (SL) blocks stacked in the ABCABC fashion, in which hexagonal atomic layers follow the Te-Bi-Te-Mn-Te-Bi-Te sequence (Fig.~\ref{fig1}a). The bonds within the blocks are ionic covalent, whereas the neighboring SLs are connected by van der Waals forces. Density functional theory (DFT) calculations coupled to the Monte Carlo simulations predict \cite{Otrokov.nat2019, Zhang.prl2019, Li.sciadv2019} that below $T_\text{Néel} = 25$ K \MBT{} develops the A-type interlayer AFM structure in which the ferromagnetically-ordered Mn layers are aligned antiparallel to each other (Fig. \ref{fig1}a). The magnetic anisotropy energy calculations revealed the easy axis with an out-of-plane orientation of the local magnetic moments of $\pm$4.6$\mu_B$. For this magnetic ground state, the insulating spectrum with the fundamental bulk bandgap of about 0.2 eV was predicted for \MBT. Having gapped spectrum and being $S$-symmetric due to the fortunate combination of its crystal and magnetic structures (Fig.~\ref{fig1}a), \MBT\, falls within the ${\mathbb{Z}}_{2}$ topological classification of AFM insulators \cite{Mong.prb2010, Fang.prb2013}. As a result of an \emph{ab initio}  calculation, ${\mathbb{Z}}_{2}=1$ was found \cite{Otrokov.nat2019, Zhang.prl2019, Li.sciadv2019} thus classifying \MBT\, as an AFM TI.

The combined $S=\Theta T_{1/2}$ symmetry protects the degeneracy of the Dirac point of the topological surface state of an AFM TI \cite{Mong.prb2010}. The surfaces that respect this symmetry are gapless, as shown in Fig. \ref{fig1}d,e for the \MBT$(10\overline{1}1)$ surface containing the $T_{1/2}$ translation. However, the $S$-breaking surfaces are gapped whenever there is a non-zero magnetization component perpendicular to them. Therefore the (0001) surface, which is \MBT's natural cleavage plane, shows a band gap of several tens of meV according to the theoretical calculation (Fig. \ref{fig1}b,c). Large predicted magnetic gap at the Dirac point is an important characteristic making this material attractive for the QAHE and TME realization. 

The magnetic and photoemission measurements performed on the bulk single crystals \cite{Otrokov.nat2019, Yan.prm2019} and molecular-beam epitaxy grown thin films \cite{Gong.cpl2019} confirm the AFM TI state in \MBT, which thus can be considered as the first \emph{intrinsic} MTI. In agreement with the theoretical predictions, a long-range AFM ordering is observed in experiment \cite{Otrokov.nat2019, Lee.prr2019, Zeugner.cm2019, Lai.prb2021}, which can be identified by the $\lambda$-like shape of the magnetic susceptibility vs. temperature curve ($\chi (T)$; Fig.~\ref{fig2}a) as well as by the presence of the characteristic spin-flop transitions in the $M(H)$ curve at $H_\text{SF}$ (Fig.~\ref{fig2}b). The N\'eel temperature $T_\text{Néel} \approx 24-26$ K is found, slightly varying from sample to sample. Subsequent neutron powder and single-crystal diffraction measurements confirmed the predicted A-type AFM ordering, as well as the out-of-plane easy axis \cite{Yan.prm2019, Ding.prb2020-2, Riberolles.prb2021}. The ARPES experiments performed on the \MBT(0001) surface revealed the bulk band gap that hosts the linearly-dispersing topological surface state with an apparent splitting at the Dirac point \cite{Otrokov.nat2019} (see Fig.~\ref{fig2}c). 

\begin{figure*}
\includegraphics[width=0.75\textwidth]{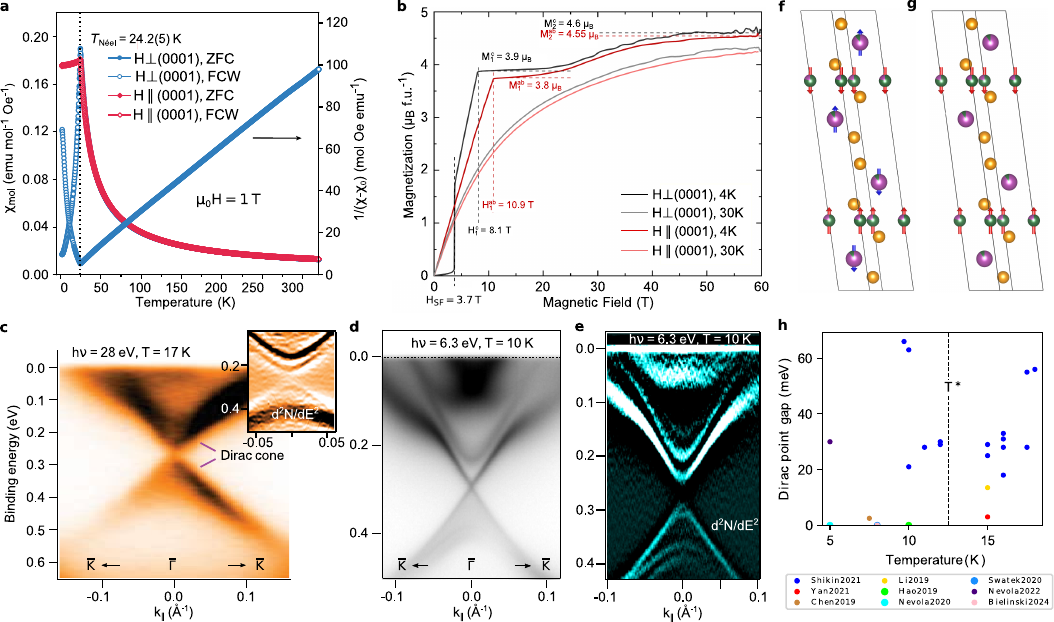}
	\caption{{\bf{Experimentally-measured magnetic properties and surface electronic structure of \MBT.}}
    {\bf a} Magnetic susceptibility (left axis) of \MBT\, as a function of temperature measured in an external magnetic field of $\mu_0H =$ 1 T in zero-field-cooled (ZFC) and field-cooled-warming (FCW) conditions, alongside the temperature-dependent reciprocal susceptibility (right axis) for $H \perp (0001)$. 
 {\bf b} High-field magnetization data of \MBT\, at 4 and 30 K with the field applied either along the $c$ axis or in the $ab$ plane. $H_\text{SF}$ denotes the spin-flop magnetic field. {\bf c, d, e} Dispersion of the topological surface state of \MBT(0001) measured with ({\bf c}) a conventional synchrotron ARPES (photon energy $h\nu = 28$ eV, $T=17$ K) and ({\bf d,e}) laser-ARPES ($h\nu = 6.3$ eV, $T=10$ K). In {\bf d,e} the data acquired on two samples synthesized by different groups are shown. In {\bf e} the second derivative data, $d^2N(E)/dE^2$, are presented. Note the reduced intensity at the expected Dirac point position in {\bf c}, indicative of the gap, which is also seen in {\bf e}, while, in contrast, the Dirac cone is gapless in {\bf d}.  In the inset to {\bf c}, the second derivative ($d^2N(E)/dE^2$) of the data taken with a more bulk sensitive photons ($h\nu = 9$ eV, $T=17$ K) is presented, where the bulk valence and conduction bands, separated by a gap of about 0.2 eV \cite{Xu.prb2021}, can be seen. {\bf f, {g}} Magnetic structure of \MBT\, below ({\bf {f}}) and above ({\bf {g}}) $T^*$, see text for the explanation. Note that the intermixing levels are intentionally enhanced for the visualization purposes. {\bf {h}}, Compilation of the literature data on the Dirac point gap size, as measured by different groups using laser-ARPES. Only the data reported for $T < T_\text{Néel}$ are shown. Note that Shikin et al. \cite{Shikin.prb2021} report data on 15 different \MBT\, samples, as shown by blue circles.  {{\bf a, c} are reproduced with permission from Ref. [\onlinecite{Otrokov.nat2019}]. 
 {\bf b} is reproduced with permission from Ref. \cite{Lai.prb2021}, Copyright American Physical Society, 2021.
 {\bf d} is reproduced from Ref. \cite{Hao.prx2019}. 
 {\bf e} is reproduced with permission from Ref. \cite{Shikin.prb2021}, Copyright American Physical Society, 2021.}}
	\label{fig2}
\end{figure*}

{{\section*{Sample-dependent Dirac point gap at \MBT(0001)}}

The above-described experimental evidences of the AFM TI state in \MBT\, were further supported by the transport experiments performed on thin films that revealed phenomena of the topological origin (see below). Very surprisingly, however, the second wave of the ARPES studies (performed with laser radiation) uncovered an unexpected behavior of the \MBT's Dirac cone, which contradicted the one expected based on the general theory of AFM TIs \cite{Mong.prb2010, Fang.prb2013}, DFT calculations made for \MBT\, \cite{Otrokov.nat2019, Zhang.prl2019, Li.sciadv2019}, and the results of the magnetic measurements performed on it. Namely, while the early photoemission studies\cite{Otrokov.nat2019, Lee.prr2019, Vidal.prb2019}, performed with the conventional synchrotron ARPES, reported large gaps in agreement with the theoretical predictions, a number of the following \emph{laser}-ARPES experiments found a dramatically reduced or even vanishingly small splitting of the \MBT\, Dirac cone \cite{Hao.prx2019, Li.prx2019, Chen.prx2019, Shikin.srep2020} (Fig. \ref{fig2}d).
Figure \ref{fig2}h summarizes all the available data on the Dirac point gap measured by laser-ARPES \cite{Yan.prb2021, Chen.prx2019, Li.prx2019, Hao.prx2019, Nevola.prl2020, Swatek.prb2020, Nevola.arxiv2022, Bielinski.arxiv2024, Shikin.prb2021}. It can be seen that in the currently available \MBT\, samples the Dirac point gap can acquire an arbitrary value between a few and several tens of meV. An example of the laser-ARPES spectrum showing a large Dirac point gap is presented in Fig. \ref{fig2}e.

Several scenarios have been theoretically proposed to account for this unexpected behavior \cite{Hou.acsn2020, Shikin.srep2020, Yuan.nl2020, Hao.prx2019, Chen.prb2021, Garrity.prm2021, Chen.prb2021}. On the one hand, possible changes of magnetism near the \MBT\, surface have been suggested, such as near-surface magnetic dead layer \cite{Yuan.nl2020} or a deviation of the magnetic structure or easy magnetization axis from those in bulk \cite{Hao.prx2019, Chen.prb2021}. Recent scanning tunneling spectroscopy study reports asymmetric quasiparticle interference pattern of the topological surface state close to the Dirac point \cite{Bian.mte2023}, which would be in line with the presence of the in-plane magnetization component at the \MBT(0001) surface. However, this is corroborated by neither the X-ray magnetic circular dichroism measurements \cite{Otrokov.nat2019, Shikin.srep2020} nor magnetic force microscopy experiments \cite{Sass.prl2020}, that provide evidences of the persistence of uniaxial A-type AFM order up to the surface layers of the \MBT\, single crystals. 

On the other hand, possible structural factors have been considered, such as surface collapse during the mechanical exfoliation \cite{Hou.acsn2020} or expansion of the first van der Waals gap \cite{Shikin.srep2020} (i.e., the increase of the spacing that separates the surface SL from the subsurface one). However, the former was only observed during the exfoliation in an inert gas environment and never in the ultra-high vacuum, while the latter needs to be quite large (about 15 \%) to close the Dirac point gap, but such expansions have not been observed in experiment \cite{Liang.prb2020, Garnica.npjqm2022}. 

{{\section*{Mn-Bi intermixing and its possible impact on the Dirac point gap}}

Latest theoretical studies discuss Mn-Bi intermixing as the most likely cause of the Dirac point gap reduction in \MBT \cite{Garnica.npjqm2022, Tan.prl2023}. To discuss the latter, let us first review the effect of the Mn-Bi intermixing on the \MBT\, magnetic structure. High-field magnetization study \cite{Lai.prb2021} revealed that above the first magnetization plateau at $\mu_0 H^\text{c}_1 \approx 8$ T (Fig. \ref{fig2}b) the magnetization smoothly increases further and eventually saturates to  $M^\text{c}_2 \approx 4.6 \mu_\text{B}$/f.u. near 50 T. The reason for such evolution of $M(H)$ from the first ($M^\text{c}_1$) to the second ($M^\text{c}_2$) plateau has been suggested a spin flip of Mn ions residing on the Bi site (Mn$_\text{Bi}$). In other words, in the ground state, each \MBT\, SL block is "ferrimagnetic" (Fig. \ref{fig2}f), in which the local moments of the Mn$_\text{Bi}$ atoms are coupled antiparallel to those of the central Mn layer, as has initially been found for \MST\, by neutron diffraction measurements \cite{Liu.prx2021}.

Going back to the discussion of possible mechanisms of the Dirac point gap reduction, Ref. [\onlinecite{Garnica.npjqm2022}] argues that it is the opposite magnetization of the Mn$_\text{Bi}$ antisites (with respect to the central Mn layer) that causes the Dirac point gap reduction. This happens because the magnetic antisites are introduced exactly in the regions of the topological surface state predominant localization in the real space, which is around the Bi layers. Thus the oppositely directed magnetic moments of the antisites can effectively counteract the magnetic effect of the central layer Mn atoms, around which the topological surface state weight is small. As illustrated by DFT calculations \cite{Garnica.npjqm2022}, even a moderate amount of the Mn$_\text{Bi}$ antisites may result in a significant reduction of the Dirac point gap or even in its almost complete shrinking. In contrast, according to the DFT results in Ref. [\onlinecite{Tan.prl2023}], the Mn-Bi intermixing "pushes" the topological surface state towards the subsurface SL, whereupon the magnetic gap is diminished heavily due to the AFM alignment between the surface and subsurface SL blocks. It should, however, be noted that both DFT studies \cite{Garnica.npjqm2022, Tan.prl2023} introduce the intermixing only in the topmost SL of the slabs used, while the other SLs are pristine. Thus, the calculations using more realistic structural models are needed to clarify which mechanism actually takes place in the experimental samples. 

Recent scanning tunneling spectroscopy studies\cite{Liu.pnas2022, Li.advmat2024} of the Dirac point gap behavior in the molecular-beam epitaxy grown thin \MBT\, films reveal the gap fluctuation across the surface. At that, the gap values ranging between 0 (gapless) and 70 meV have been measured. These studies differ, however, as far as the correlation of the gap size with the point defects distribution is concerned. While Ref. [\onlinecite{Liu.pnas2022}] reveals that the gap is absent (present) in the regions with a high (low) concentration of Mn$_\text{Bi}$ defects, Ref. [\onlinecite{Li.advmat2024}] concludes that the gap size is uncorrelated to individual Mn$_\text{Bi}$ and Bi$_\text{Mn}$ defects. Instead, the Dirac point gap fluctuation appears to take place on the nanometer scale. Remarkably, application of the external magnetic field of about 1 T significantly reduces the gap fluctuations and leads to the increase of the average gap from 26.3 meV to 44 meV \cite{Li.advmat2024}. The sensitivity of the Dirac point gap to the moderate magnetic field could be consistent with a deviation of the near-surface magnetic structures from the bulk one, as mentioned above \cite{Bian.mte2023}.

In this context, an important comment should be made. Out of all defects generated by the Mn-Bi intermixing, only the Mn$_\text{Bi}$ antisites from the first Bi layer (i.e., the second atomic layer from the surface) are clearly seen in the scanning tunneling microscopy. Imaging of the Bi$_\text{Mn}$ defects that lie in the fourth atomic layer from the surface is already quite difficult \cite{Huang.prm2020, Garnica.npjqm2022, Li.advmat2024}. As far as the the sixth atomic layer Mn$_\text{Bi}$ defects are concerned, to the best of our knowledge, no feature that would be clearly attributed to them has been reported in the scanning tunneling microscopy studies. These features should undoubtedly be faint and, because their extension should be about several lattice parameters as a minimum, may laterally overlap with each other as well as with those from defects in other layers, making them hardly distinguishable. Nevertheless imaging of the sixth layer Mn$_\text{Bi}$ antisites is highly desirable since, according to the DFT calculations\cite{Garnica.npjqm2022}, their negative effect on the Dirac point gap is expected to be much stronger than that of the second layer Mn$_\text{Bi}$. This is because the sixth atomic layer (again, counting from the surface) carries a much larger weight of the topological surface state in the real space\cite{Garnica.npjqm2022}.

Latest experimental data shed more light onto the magnetism of the \MBT\, family, which might be relevant in the context of the Dirac point gap issue. The nuclear magnetic resonance and muon spin spectroscopy measurements not only directly confirm the ferrimagnetic structure of the \MBT\, SLs (and those of \MBT$\cdot n$\BT\, discussed below), but also reveal that the static magnetic moment of the Mn$_\text{Bi}$ antisite sublattice disappears at $T^*<T_\text{Néel}$, $T^*$ being equal to 12.5 K in the \MBT\, case \cite{Sahoo.advsci2024}. I.e., below (above) $T^*$ the Mn$_\text{Bi}$ sublattice is ordered (paramagnetic), as shown in Fig. \ref{fig2}f({g}). It would be very interesting to explore the Dirac point gap behavior across $T^*$, given that one of the proposed scenarios\cite{Garnica.npjqm2022} attributes the gap reduction to the AFM coupling between the Mn$_\text{Bi}$ and Mn$_\text{Mn}$ sublattices when both of them are ordered (i.e., below $T^*$). The existing temperature-dependent photoemission data \cite{Shikin.prb2021} do not elucidate this behavior since the lowest measurement temperature {($\sim 10 $ K)} is too close to the expected $T^*$. {Note that for the samples studied in \cite{Shikin.prb2021} $T^*$ are unknown. It is reasonable to assume that $T^*$ can slightly vary from sample to sample (likely depending on the degree of intermixing), pretty much as the Néel temperature of \MBT\, varies within the 24-26 K range \cite{Garnica.npjqm2022, Sahoo.advsci2024}. Taking also into account the instrument error in the temperature determination and the temperature step of about 1.7 K in Ref. \cite{Shikin.prb2021}, it is difficult to draw conclusions about the Dirac point gap behavior across $T^*$ from that experiment.} The compilation of the laser-ARPES results shown in Fig. \ref{fig2}{h} does not clarify this issue either, since the measurements have been performed by different groups on different samples.
Meanwhile, using time- and angle-resolved photoemission spectroscopy, it has been found that opposite helicities of mid-infrared circularly polarized light result in substantially different Dirac mass gaps in the \MBT\, topological surface state below $T^*$ (at 8 K), despite the equilibrium Dirac cone being massless \cite{Bielinski.arxiv2024}.

{\section*{\MBT\, surface electronic structure above $T_\text{Néel}$} }

At this point, the surface electronic structure of \MBT\, above the Néel temperature should also be discussed. A number of ARPES studies, both synchrotron \cite{Otrokov.nat2019, Lee.prr2019, Vidal.prb2019} and laser based \cite{Otrokov.nat2019, Li.prx2019, Estyunin.aplmat2020, Shikin.srep2020, Shikin.prb2021}, in which the Dirac point gap has been detected in the AFM state, reported its persistence in the paramagnetic (PM) phase, similarly to what was previously observed for the magnetically-doped \BS\, \cite{Chen.sci2010, Xu.nphys2012}. At that, in majority of the cases, \MBT's Dirac point gap size in the AFM and PM phases is practically the same, with only one laser-ARPES study reporting a $\sim$40\% reduction of the gap (from 65 to 40 meV) upon heating from below $T_\text{Néel}$ up to 35 K \cite{Shikin.prb2021}. 
Such a behavior challenges the idea of the magnetically induced gap, as the latter should disappear at or just above the critical point, pretty much as it happens with the exchange splitting of the first bulk conduction and valence bands observed for \MBT\, \cite{Chen.prx2019, Estyunin.aplmat2020}. However, in \MBT\, the persistence of the Dirac point gap could be caused by strong short range order effects that exist up to about 60 K, as observed by electron spin resonance, as well as ferromagnetic and antiferromagnetic resonance experiments \cite{Otrokov.nat2019, Alfonsov.prb2021, Alfonsov.prb2021-2}. The measured magnetization data \cite{Lai.prb2021}, revealing that \MBT\, is not in the PM limit even at $T \approx 50$ K, support the latter observations. Such a behavior is also consistent with the strong spin fluctuation-driven spin scattering above $T_\text{Néel}$ found in a previous magneto-transport study of \MBT\, in Ref. [\onlinecite{Lee.prr2019}]. 
Beyond 50-60 K, an unprecedentedly large anisotropy of the Mn spin relaxation rate in the PM state of \MBT\, \cite{Otrokov.nat2019, Alfonsov.prb2021} may give rise to an instantaneous (on the timescale of electron spin resonance) out-of-plane magnetic field at the surface, preventing the gap from closing \emph{on the much faster timescale of the ARPES experiment}.

{\section*{MTIs of the \MBT\, family}}

In spite of the Dirac point gap puzzle, \MBT\, has attracted significant interest of the research community due to its unusual and highly tunable properties. Multiple tuning knobs, not only extrinsic such as magnetic field, pressure, and temperature, but also intrinsic such as Mn-Mn interlayer distance, chemical composition, and defect engineering, can be used to realize various magnetic and topological states in \MBT\, or systems on its basis. For example, pnictogen or chalcogen substitutions and Mn/Bi/Te stoichiometry alternations give rise to  such materials as \MBST, \MST, \MBS, or Mn$_2$Bi$_2$Te$_5$, whose magnetic and topological properties were studied both theoretically and experimentally \cite{Yan.prb2019, Chowdhury.npjcm2019, Wimmer.advmat2021, Lee.prx2021, Zhu.nl2021, Cao.prb2021, Eremeev.prb2022, Watanabe.apl2022, Lupke.cmat2023}. 
In particular, \MBS, while being AFM TI with the same A-type structure as \MBT, displays staggered magnetization within the Mn plane \cite{Zhu.nl2021}, which may give rise to the one-dimensional topologically-protected flat bands pinned to the magnetic domain walls at its surface, as predicted by theory \cite{Petrov.prb2021}. Incidentally, recent experimental and theoretical X-ray magnetic circular dichroism study has revealed that the Mn$_\text{Bi}$ antisites in the SLs of the \MBS/\BS\, heterostructure are coupled ferromagnetically to the central Mn plane of the SL \cite{Fukushima.prm2024}, in a stark contrast to the \MBT\, case, where this coupling is antiferromagnetic \cite{Lai.prb2021, Sahoo.advsci2024}. This striking difference has been attributed to a stronger hybridization of the Mn$_\text{Bi}$-$d$ states with the Se-$p$ states in \MBS\, as compared to that with Te-$p$ in \MBT, due to the shorter bondlength in the former. Another material with the 1-2-4 stoichiometry, \MST, seems to be on the verge of the topological phase transition, as the DFT calculations reveal a strong sensitivity of its topology to the crystal structure details \cite{Eremeev.jpcl2021, Zhang.prl2019, Chen.ncomms2019, Lei.pnas2020, Liu.prx2021}, while the available experimental data are contradictory \cite{Wimmer.advmat2021,Xi.jpcl2022} and further studies are needed. Mixing Bi and Sb on the pnictogen sublattice to create \MBST, enables realization of the Weyl semimetal state in the fully polarized ferromagnetic (FM) state \cite{Zhang.prl2019, Li.sciadv2019}, which was experimentally achieved by applying a sufficiently strong external magnetic field that overcomes the AFM interlayer exchange coupling \cite{Lee.prx2021,Chong.ncomms2023}.
Ge-, Sn-, and Pb-doped \MBT\, single crystals were also synthesized \cite{Zhu.prb2021,Qian.prb2022,Estyunin.mdpi2023}, showing peculiar topological phase transition \cite{Frolov.commphys2024}. 

Furthermore, the van der Waals nature of \MBT\, enables "intercalating" the adjacent SLs with \BT\, quintuple layers, resulting in the \MBT·$n$\BT\, family of compounds ($n=1$ for \MBTBT,  $n=2$ for \MBTBTBT, and so on, up to $n=6$)~\cite{Aliev.jac2019, Souchay.jmcc2019, Jahangirli.jvstb2019, Amiraslanov.prb2022}. The increasing distance between the SLs progressively weakens the interlayer exchange coupling with an increasing $n$ \cite{Zverev.jetp2023}, which enables an effective tuning of the magnetic structure by moderate magnetic fields~\cite{Vidal.prx2019, Wu.sciadv2019, Hu.ncomms2020, Klimovskikh.npj2020, Wu.prx2020, Lu.prx2021}, or hydrostatic pressure~\cite{Shao.nl2021}, driving these compounds from the AFM to the FM state. Overall, apart from the 3D AFM TI state, under different conditions \MBT·$n$\BT\, materials may show topological crystalline insulator state \cite{Vidal.prx2019} as well as higher order topologies such as M{\"o}bius \cite{Zhang.prl2020} or ${\mathbb{Z}}_{4}$-axion \cite{Zhang.prl2020, Hu.sciadv2020, Tanaka.prr2020, Gu.ncomms2021} insulators, 2D AFM second order TI\cite{Zhan.nl2024} or topological superconductor\cite{Roy.prb2020} states. 

Magnetic antisite defects in the materials of the \MBT\ family are lately in the center of attention not only because of their possible negative impact on the Dirac point gap but also thanks to their strong influence on the magnetic and electronic structure: they are exploited as an effective tuning knob to purposely modify the latter~\cite{Liu.prx2021, Tcakaev.advsci2023, Yan.nl2022}. The cation intermixing between the manganese and pnictogen crystallographic sites is favored by closeness of their ionic radii. This is especially true for \MST\, that supports up to $\sim$40\% of Sb atoms in the Mn layer, while, in turn, up to $\sim$15\% of each Sb layer is occupied by Mn atoms ~\cite{Liu.prx2021}. For \MBT\, the intermixing levels are roughly three times lower \cite{Zeugner.cm2019}. Strong intermixing in \MST\, promotes the FM coupling between the adjacent SLs\cite{Murakami.prb2019}. At that, the magnetic transition temperature jumps from $T_\text{Néel} = 19$ K in the AFM-like bulk \MST\, single crystals~\cite{Liu.prx2021} to {$T_\text{Curie}=73$ K} in the FM-like ones~\cite{Kochetkova.cm2025}, which is achieved by varying the growth conditions with a certain degree of control \cite{Liu.prx2021}. According to the DFT calculations, while a moderate Mn-Sb intermixing promotes the Weyl semimetal state due to the change of the interblock coupling to FM, its increase eventually renders \MST\, gapped and topologically trivial \cite{Wimmer.advmat2021, Liu.prx2021}. Apart from \MST, the interlayer coupling can also become truly FM in the Sb-doped \MBT$\cdot n$\BT\, \cite{Bo.prb2021, Xie.jpd2021} and even in \MBTBTBT, in the latter case via the Mn-Bi defects engineering under appropriate growth conditions~\cite{Tcakaev.advsci2023, Yan.nl2022}, which may help to achieve an FM axion insulator state~\cite{Zhang.prl2020, Hu.sciadv2020}. 

{{\section*{\MBT\ in the two-dimensional limit}}

\MBT\, properties in the two-dimensional limit are exciting, too. Magnetism and topology of thin \MBT\, films are thickness-dependent, while different symmetries in even- and odd-SL films lead to distinct phenomena \cite{Li.sciadv2019, Otrokov.prl2019, Gong.cpl2019, Trang2021:ACSN, Zhao.nl2021}. The films with odd (even) number of SLs are uncompensated (fully compensated) interlayer antiferromagnets, which break (preserve) the $P\Theta$ symmetry \cite{Li.sciadv2019}, $P$ being inversion. In the films made of even number of blocks, the $P\Theta$ symmetry leads to the Chern number $C = 0$, while a $C \neq 0$ is allowed in the films with the odd number. According to DFT calculations \cite{Li.sciadv2019, Otrokov.prl2019}, while the one-SL-thick film is topologically trivial, the 5- and 7-SL-thick ones are QAH insulators with $C = 1$ {(the predictions on the 3-SL-thick film are contradictory \cite{Li.sciadv2019, Otrokov.prl2019})}. On the other hand, the 2-, 4-, and 6-SL-thick films were predicted to show the so-called zero plateau QAH state. The importance of the zero plateau QAH state is that it can be a suitable platform for realization of the TME, characteristic of the axion insulator phase \cite{Wang.prb2015}. 

Soon after the theoretical predictions \cite{Li.sciadv2019, Otrokov.prl2019}, the  zero plateau QAH state has been observed experimentally in the \MBT\, thin flakes made of \emph{even} number of SLs \cite{Liu.nmat2020}, see Fig. \ref{fig3}a. Previously, this  state of matter was being sought for in the FM1/TI/FM2 QAH heterostructures, where a relatively thick TI spacer enables magnetization reversal of the individual FM layers that have different coercivities, leading to the overall AFM alignment and, consequently, to a zero plateau QAH state \cite{Mogi.sciadv2017, Xiao.prl2018}. In theory, the \MBT\, thin films made of even number of SLs realize this state intrinsically, i.e., without the need of magnetic field application. In practice, however, a magnetic field of about 8 T is applied and swiped to 0 T to prepare a single-domain AFM state with the desired orientation of the topmost SL magnetization, up or down \cite{Gao.nat2021, Gao.natel2024}. It should also be noted that a thickness of 6 SLs \cite{Liu.nmat2020} is likely insufficient for the TME to be quantized because of the finite-size effect \cite{Wang.prb2015, Fijalkowski.prb2021, Zhuo.ncomms2023}, however it should become fully quantized upon increasing the film thicknesses towards the 3D limit.

Further experimental work not only confirms the zero plateau QAH state in 2D even-layered \MBT, \cite{Gao.nat2021, Qiu.nmat2023} but also reveals a series of other novel phenomena \cite{Gao.nat2021, Qiu.nmat2023, Gao.sci2023, Wang.nat2023, Gao.natel2024}. One of them is the layer Hall effect, in which the electrons from the top and bottom layers of the film spontaneously deflect in the opposite directions, which is due to the opposite signs of the Berry curvature \cite{Gao.nat2021, Chen.nsr2024}. Observation of the effect is enabled by applying external electric field that breaks $P\Theta$, leading to the emergence of a finite {anomalous Hall effect} {(}AHE{)}, which is dominated by either top or bottom layer, depending on the field direction. Moreover, the opposite sign of the Berry curvature of the top and bottom surfaces is responsible for the appearance of the axion {(i.e., surface \cite{Malashevich.prb2010})} contribution to the optical magneto-electric coupling, as recently predicted \cite{Ahn.ncomms2022} and observed for the 2D even-layered \MBT\, \cite{Qiu.nmat2023}.  This effect enables optical control of the AFM order, i.e., selective stabilization of one or another AFM domain (up-down or down-up) by the circularly polarized light with either opposite helicity or different frequencies that couples differently to the opposite AFM domains.

Another unusual phenomenon recently witnessed in the thin \MBT\, flakes with even number of blocks is the second-order nonlinear AHE \cite{Gao.sci2023,Wang.nat2023}, in which an alternating current (a.c.) with frequency $\omega$ induces an a.c. Hall voltage with frequency $2\omega$. In this case, the Berry curvature is not the origin of the effect, since it is equal to zero as dictated by the $P\Theta$ symmetry. Instead, the effect is generated by the quantum metric of the gapped \MBT\, Dirac cone \cite{Gao.sci2023, Wang.nat2023}. More recently, third-order nonlinear Hall effect in both longitudinal and transverse directions in thick \MBT\, flakes has been reported \cite{Li.ncomms2024-2}. The longitudinal third-harmonic response $V^{3\omega}_{xx}$ has been found to appear due to the quantum  metric quadrupole, while the transverse response $V^{3\omega}_{xy}$ due to the Berry curvature quadrupole. Thus, \MBT\, enables probing both the real and the imaginary part of the quantum geometric tensor of the Bloch states in transport experiments.

As for the predicted QAHE (i.e., the {quantum Hall effect} {(}QHE{)} in zero magnetic field), 
its observation in the \emph{odd-layered} 2D \MBT\, has proven challenging. In fact, initially \MBT\, has shown the quantized Hall effect only under external magnetic field \cite{Deng.arxiv2019, Liu.arxiv2019}, but not without it {(note that Refs. \cite{Deng.arxiv2019, Liu.arxiv2019} are preprint versions of \cite{Deng.sci2020, Liu.nmat2020})}. The field has been used to overcome the AFM interlayer exchange coupling thus forcing the FM state (Fig. \ref{fig3}a), while the gate bias is employed to find the charge neutrality point, which in \MBT\, should correspond to the gap in the Dirac point. {At the magnetic fields used (about 9 T), the magnetic moments of the Mn$_\text{Bi}$ antisites are most likely still ordered oppositely to those of the central Mn layer \cite{Lai.prb2021}. However, this does not suppress the quantized Hall effect.}
{In fact, the effect can be observed already in the canted AFM state, just after the spin-flop transition \cite{Cai.ncomms2022}. Upon further increase of the magnetic field strength, the canted AFM state transforms into the forced FM state with the QHE observed continuously.} The absence of the free carriers at the Fermi level translates into the absence of Landau levels (LLs) in spite of the field application (in contrast to the conventional QHE \cite{vonKlitzing.prl1980}) and the QHE stems from the non-trivial topology of the \MBT's band structure in the forced FM state.  Envisioned by theory \cite{Li.sciadv2019, Otrokov.prl2019}, this $C=1$ QHE without LL formation in \MBT\, has been repeatedly observed in experiment by different groups to this date \cite{Deng.sci2020, Liu.nmat2020, Ge.nsr2019, Gao.nat2021, Ovchinnikov.nl2021, Hu.prm2021, Liu.ncomms2021, Ying.prb2022, Cai.ncomms2022, Bai.nsr2023}, with a reasonable quantization persisting up to 30 K \cite{Ge.nsr2019}. Worth mentioning as well is the $C=2$ state without LLs that onsets in the flakes thicker than 9 SLs of \MBT\, under external magnetic field and persists up to about 10 K \cite{Ge.nsr2019,Lei.pnas2020}. Finally, the conventional QHE \emph{with} LLs \cite{vonKlitzing.prl1980} can be observed in the \MBT\, flakes too  \cite{Deng.sci2020}, which, apart from a rather strong magnetic field, requires increasing the carrier concentration by applying bias voltage.

The zero-field QAHE has been observed in the 5-SL-thick film \cite{Deng.sci2020}, as shown in Fig. \ref{fig3}c,d. However, further advances in this direction have been slow, likely due to the difficulties in obtaining high-quality samples with reduced levels of the Mn-Bi intermixing, although the device fabrication process seems to introduce further complications \cite{Hou.acsn2020, Li.ncomms2024}. A valuable insight into the quantized transport of the odd-layered \MBT\, flakes has been provided by scanning superconducting quantum interference device microscopy used to image their current distribution \cite{Zhu.arxiv2023}. A chiral edge current in the 7-SL-thick flake at zero magnetic field has been observed, confirming its topological nature. However, the finite bulk conduction and edge-bulk scattering have been found to undermine the transport quantization. This is consistent with the scanning tunneling spectroscopy data that image both the edge state of the 5-SL-thick molecular-beam epitaxy grown \MBT\, film and its coupling to the gapless two-dimensional bulk regions arising from band gap ﬂuctuations \cite{Li.advmat2024}. However, very recently, a new procedure in the device preparation, consisting in depositing an AlO$_x$ layer on the \MBT\, flake surface prior to nano-fabrication, has enabled a significant improvement of the AHE phase quality, with the true zero-field QAHE achieved in few samples \cite{Wang.arxiv2024, Lian.arxi2024}. Applying a hydrostatic pressure also allowed approaching the QAHE regime in 2D \MBT\,  \cite{Chong.arxiv2023}. 

\begin{figure}
 \includegraphics[width=0.49\textwidth]{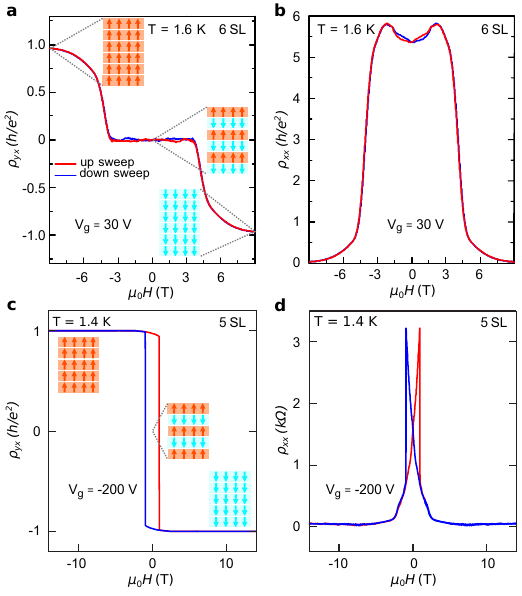}
	\caption{{\bf{Q(A)HE and its zero-plateau in \MBT{} flakes.}}
    {\bf a,b} Magnetic field driven transition from the zero-plateau QAH state (around the zero field, corresponding to $\rho_{yx}=0$ ({\bf a}) and very large $\rho_{xx}$ ({\bf b})) to the Chern insulator with QHE (at $\pm9$ T, with the nearly quantized $\rho_{yx}\approx \mp h/$e$^2$ ({\bf a}) and vanishing $\rho_{xx}$ ({\bf b})) measured in a 6-SL-thick \MBT\, flake at $T=1.6$ K. The measurement is performed using the back-gate voltage $V_g = 30$ V, which corresponds to the charge neutrality point when the Fermi level lies within the surface state gap. {\bf c,d}  Quantum anomalous Hall effect in a 5-SL-thick \MBT\, flake illustrated by the magnetic field dependence of $\rho_{yx}$ ({\bf c}) and $\rho_{xx}$ ({\bf d}), measured at $T = 1.4$ K and $V_g = -200$ V. At $\mu_0 H=0$ T, $\rho_{yx}$ reaches 0.97$h/$e$^2$, while $\rho_{xx}$ is suppressed (0.061$h/$e$^2$), evidencing the QAHE. The insets in panels {\bf a,c} schematically show the magnetic structure of the flakes at zero and high fields. {{\bf a, b} are reproduced with permission from Ref. \cite{Liu.nmat2020}, Copyright Springer-Nature, 2020. {\bf c, d} are reproduced with permission from Ref. \cite{Deng.sci2020}, Copyright AAAS, 2020.} }
	\label{fig3}
\end{figure}

Recently, an intriguing behavior has been reported for the molecular-beam epitaxy grown 5 SL \MBT\, film showing the QHE without LLs in the forced FM state. Tuning the gate voltage through the QHE region up until the quantization breakdown, when $\rho_{yx}$ is no longer equal to $h/$e$^2$, and then switching the field off (for the material to adopt the AFM state) leads to reappearance of a nearly quantized $\rho_{yx}$ plateau of $\approx$$h/$e$^2$ \cite{Li.arxiv2024}. This happens for the Fermi level located in the valence band and not in the global gap. Despite the latter, as well as nonzero $\rho_{xx}\approx 0.85 h/$e$^2$, this state shows tan($\theta_\text{H}$) vs. $T$ behavior ($\theta_\text{H} = \rho_{yx}/\rho_{xx}$ is the Hall angle), which is more similar to that of the zero-field QAHE in \MBT\, \cite{Deng.sci2020, Chong.arxiv2023}, rather than the insulating zero field state of the very same epitaxial film at the charge neutrality point. The appearance of this phase, dubbed "reentrant QAHE", has been attributed to the exchange-induced Berry curvature splitting and disorder-induced Anderson localization \cite{Li.arxiv2024, Chen.scp2021}.

{{\section*{Outlook}}

{Let us now briefly outline other exciting phenomena that compounds of the \MBT\ family have been reported to (or may potentially) host. While we do not delve into details -- an undoubtedly worthwhile endeavor that lies beyond the scope of this short review -- we nevertheless highlight these intriguing studies as a roadmap for future research.}
Apart from the above-described variants of the Hall effect, the compounds of the \MBT-family were reported to show the QAHE in the 3D limit \cite{Deng.nphys2021}, as well as the topological \cite{Roychowdhury.chemmat2021, Takashiro.nl2022} and planar Hall effects \cite{Wu.nl2021}. Pending the observation are the predicted half-integer QHE \cite{Mong.prb2010}, $\Theta$-breaking quantum spin Hall effect \cite{Sun.prl2019} and its "hinged" version \cite{Ding.prb2020}, interfacial crystal Hall effect \cite{Shao.pra2021}, orbital Hall effect \cite{Chen.nl2024}, as well the quantized version of the layer Hall effect \cite{Dai.prb2022}.
Besides, \MBT-based systems could be a platform for novel topological heterostructures\cite{Hirahara.nl2017, Otrokov.2dmat2017, Sun.prl2019, Hirahara.ncomms2020, Kagerer.prr2023, Fukasawa.prb2021, Li.advmat2022, Klimovskikh.mta2024}, anomalous Nernst effect \cite{Ceccardi.npjqm2023}, high-Chern number phases \cite{Bosnar.npj2d2023}, skyrmions \cite{Jiang.apl2021} (in particular, coexisting with QAHE, which may generate novel topological phases \cite{Li.pnas2022}), and Majorana fermions \cite{Peng.prb2019}. Recent studies are seeking to induce superconductivity in \MBT\, either via proximitizing it to a conventional superconductor NbSe$_2$ \cite{Dong.arxiv2023} or via interfacing \MBT\, film to that of a non-superconducting antiferromagnetic material FeTe \cite{Yuan.nl2024}.

Novel device proposals based on the systems of the \MBT-family have already been put forward \cite{Perez-Piskunow.prl2021, An.npjcm2021}, such as rectifiers, spin filters, negative differential resistive devices, photoelectric sensors, photovoltaic, magneto-optoelectronic devices. The layer Hall effect makes \MBT\, a potential platform to explore the ‘layertronics’ to encode, process and store information \cite{Li.nsr2024}, while the recently observed AFM diode effect \cite{Wang.nat2023, Gao.natel2024} may enable a field-effect transistor and harvesting of wireless electromagnetic energy \cite{Gao.natel2024}. Quantum computing and sensing applications have also been envisioned \cite{Varnava.ncomms2021, Pixley.patent2022}.

It should be said, however, that this research field is only in its infancy and many results are still to be confirmed and understood, while many new exciting discoveries are likely to come soon. The most important task appears to be achieving an exhaustive explanation of the unexpected behavior of the \MBT\, topological surface state gap and further detailed studies are needed. However, since the Mn-Bi intermixing seems to be detrimental in this sense, suppressing it could hopefully allow getting rid of the Dirac point gap issue in \MBT. Steps in this direction are being taken currently \cite{Hu.prm2021, Yan.jac2022, Liu.pnas2022, Bai.nsr2023, Hu.nsr2024}. Achieving this objective might translate into the improved quality of the experimental samples and, hence, to an observation of a plethora of phenomena and states of matter the \MBT\, family is capable of hosting. In particular, it could enable realization of the robust intrinsic QAHE in \MBT, whose observation temperature may be expected to jump significantly above the currently reported 1.5 K \cite{Deng.sci2020}.

For further detailed reading on the intrinsic MTIs of the \MBT-family we recommend the following review articles \cite{Wang.inn2021, Jiang.es2023, Li.nsr2024review, Hu.nsr2024}.

{We have recently learned of the experimental observation of the topological Anderson Chern insulator phase in the [\MBT]$_\text{1SL}$/[\BT]$_\text{1QL}$ heterostructure (where QL stands for quintuple layer), as reported in \cite{Wang.arxiv2025}.}

\section*{ACKNOWLEDGMENTS}
We thank Dr. I.I. Klimovskikh for stimulating discussions. We acknowledge the support by MCIN/AEI/10.13039/501100011033/ (Grant PID2022-138210NB-I00) and "ERDF A way of making Europe", by the Grant CEX2023-001286-S funded by MICIU/AEI/10.13039/501100011033, as well as MCIN with funding from European Union NextGenerationEU (PRTR-C17.I1) promoted by the Government of Aragon. A.Yu.V. acknowledges
the Ministry of Science and Higher Education of the Russian Federation (state task No FSWM-2025-0009). E.V.C. acknowledges Saint-Petersburg State University for a research project No 116812735. {M.B. acknowledges support of the Croatian Science Foundation under the project numbers HRZZ-IP-2022-10-6321 and HRZZ-MOBDOL-2023-12-6938.}

\section*{Author contributions}
M.M.O. and E.V.C. conceived the idea and supervised the project. A.Yu.V., M.B., and M.M.O. made the literature analysis. A.Yu.V. and M.B. prepared the illustrations. All authors contributed to the writing of the manuscript.

\section*{Competing interests}
The authors declare no competing interests.


\end{document}